\documentclass[submitted]{IEEEtran}
\usepackage{cite}
\usepackage{amsmath}
\usepackage{graphicx}
\usepackage{amssymb}
\usepackage{upgreek}
\usepackage{multirow}
\usepackage{multicol}
\usepackage{tabularx}
\usepackage[colorinlistoftodos]{todonotes}
\usepackage[breaklinks=true,colorlinks=true,linkcolor=blue,urlcolor=blue,citecolor=blue]{hyperref}
\usepackage[colorinlistoftodos]{todonotes}

\begin{document}

\title{An Approach for Restoring Magnetic Field Uniformity in Openable BIPM-Type \\Kibble Balance Magnets }
        \author{Nanjia Li, Weibo Liu, Yongchao Ma, Wei Zhao, Songling Huang, {\it Senior Member, IEEE},\\ Shisong Li$^\dagger$, {\it Senior Member, IEEE}
        \thanks{Nanjia Li, Weibo Liu, Yongchao Ma, Wei Zhao, Songling Huang, and Shisong Li are with the Department of Electrical Engineering, Tsinghua University, Beijing 100084, China. Wei Zhao is also with the Yangtze Delta Region Institute of Tsinghua University, Jiaxing, Zhejiang 314006, China. $^\dagger$Email: shisongli@tsinghua.edu.cn.}
} 

\markboth{}{}
\maketitle

\begin{abstract}
The Kibble balance realizes the kilogram by linking mechanical and electrical quantities via a magnet system. In an improved BIPM-type magnet design by Tsinghua University, an open/close surface was incorporated, facilitating operation. However, an unavoidable mechanical air gap at the splitting plane introduces asymmetry in the magnetic flux density profile, degrading field uniformity. This study proposes a two-step yoke compensation method to restore symmetry by adjusting the upper outer yoke's inner radius and the splitting gap height. Finite element simulations show linear relationships between asymmetry and these parameters, enabling predictive compensation. Experimental results confirm that sequential tuning successfully eliminates asymmetry and recovers the designed uniform field range. The method provides an effective solution for enhancing magnetic field quality in openable Kibble balance magnets.
\end{abstract}


\section{Introduction}

\IEEEPARstart{T}{he} Kibble balance is a primary instrument for realizing the kilogram with a relative uncertainty of $10^{-8}$~\cite{kibble1976measurement}. It precisely links mechanical and electrical power via two phases: weighing and velocity. In the weighing phase, a current $I$ through a coil in a magnetic field produces an electromagnetic force balancing the gravitational force on a test mass $m$, given by $mg = BlI$, where $g$ is gravitational acceleration, $B$ is magnetic flux density, and $l$ is the coil's effective length. In the velocity phase, the coil moves vertically at velocity $v$, inducing a voltage $U = Blv$. Combining these yields $m = UI/(gv)$. Since $U$, $I$, $g$, and $v$ are measurable via quantum electrical standards, optical interferometry, and frequency standards, the Kibble balance ultimately relates the kilogram to the Planck constant $h$~\cite{haddad2016bridging}. For further details, see~\cite{Robinson_2016}. 

Although the geometric factor $Bl$ does not explicitly appear in the final mass equation, it serves as the essential link between the two measurement phases. Nearly all Kibble balances currently use yoke-based permanent magnet systems, as reported in~\cite{Wood_2017,NIST2017,METAS2022,Fang_2020BIPM,NIM2023,KRISS2020,UME2023,MSL2020,NPLtabletop2024,LNE2025,THU2025,Rothleitner_2018}. Among these, the magnet circuit originally developed by the  Bureau International des Poids et Mesures (BIPM) group has become the most widely adopted design~\cite{BIPMmag,NISTmag,NIMmag,METAS_Baumann_2013,Kim_KRISS_2014,THUmag,UME_Ahmedov_2018,Rothleitner_2018}, due to advantages such as high field strength, compact size, and excellent self-shielding. Further details on magnet designs are available in the recent review~\cite{Li_2022}.

\begin{figure*}[t]
\centering
\includegraphics[width=0.9\textwidth]{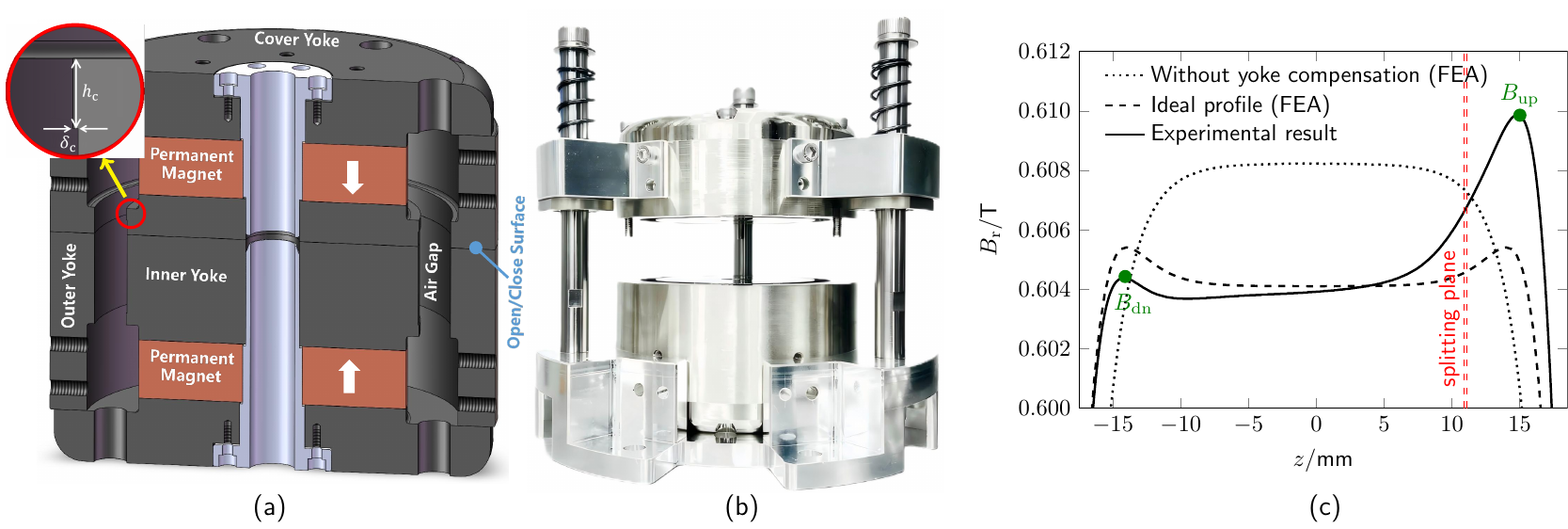}
\caption{(a) Schematic of the modified BIPM magnet: permanent magnets (red), yoke (gray), and splitting plane (blue). Arrows indicate magnetization. The red circle highlights the inner yoke compensation structure (height $h_\mathrm{c}$, width $\delta_\mathrm{c}$) for improved field uniformity. (b) Photo of the magnet in open state. Splitting enables levitation and yields 92\,N attraction when closed. (c) Magnetic field comparison: dotted curve (conventional design), dashed curve (with yoke compensation), solid curve (experimental result). The splitting gap causes a 5.5\,mT intensification ($\Delta B_\mathrm{r}=B_{\rm{up}}-B_{\rm{dn}}$) above the plane, reducing uniformity. Red dashed lines indicate the open/close plane.}
\label{fig01}
\end{figure*}

The Tsinghua Kibble balance group has enhanced the BIPM-type magnet system through two innovations~\cite{THUmag}: 
1) As illustrated in Fig.~\ref{fig01}(a), an inner yoke compensation structure, initially predicted by finite element analysis (FEA)~\cite{9042232}, is incorporated into the conventional BIPM design, significantly expanding the uniform field range $B_\mathrm{r}(z)$ as experimentally validated. {The idea is narrowing the gap width at both ends, which concentrates a greater percentage of the magnetic flux there, strengthening the field. This increase compensates for the edge effect, thereby enlarging the region of field uniformity.} 
2) An optimized open/close plane is introduced as shown in Fig.~\ref{fig01}(b), enabling magnetic levitation in the open state and easy separation when closed, greatly simplifying magnet operation. However, the small unavoidable air gap between components compromises magnetic symmetry and introduces vertical field gradients. An experimental profile presented in Fig.~\ref{fig01}(c) shows asymmetric flux density with higher values near the splitting plane, reducing field uniformity.
{Note that in Kibble balances, enlarging the field uniform range is beneficial, including: 1) It allows for a larger coil winding space and enables a coil design with an optimal \( Bl \) value~\cite{NIST4magdesign} while maintaining lower resistance, which in turn mitigates ohmic heating and reduces the associated thermal-magnetic effects~\cite{Magneticerror,thermalerror}. 2) A uniform magnetic field---characterized by a quasi-one-dimensional profile following a \( 1/r \) distribution---renders the \( Bl \) value insensitive to parasitic motions of the coil~\cite{flatprofile}, thereby minimizing misalignment errors. 3) When the coil velocity is constant, a uniform field produces a flatter induced voltage profile, which simplifies integration with differential voltage measurements using a DC reference. 4) A larger uniform field range extends the usable range for \( Bl \) measurements, which is essential for reducing fitting errors during data analysis.
A common technique for reducing field gradients is to reshape the air gap-yoke boundary, e.g.~\cite{BNMmag,METASmag}. However, for precision adjustments, the required surface modifications can be on the micrometer scale. At this level, yoke nonlinearities can cause the realized field profile to deviate significantly from the target~\cite{NISTmag}.} To address this issue, in this paper, we propose a two-step yoke structure compensation method that enables flexible adjustment of the magnetic profile flatness. The principle of this approach is presented in section~\ref{sec02}, followed by experimental validation in section~\ref{sec03}. The work concludes in section~\ref{sec04}.

\section{Principle of the proposed magnetic profile optimization method}
\label{sec02}

First, the cause of the magnetic profile asymmetry, as shown in Fig. \ref{fig01}(c), is {illustrated}. The analysis uses a magnet system similar to the Tsinghua design~\cite{THUmag}, comprising two NdFeB permanent magnet disks (inner radius 10\,mm, outer radius 45\,mm, height 20\,mm). Their flux is guided through a 11\,mm wide, 40\,mm high air gap by high-permeability yokes, producing a flux density of approximately 0.6\,T at the mean radius. {As shown in the zoom area of Fig. \ref{fig01}(a), a ring compensation structure with a rectangular cross-section} ($\delta_\mathrm{c} = 0.3$\,mm, $h_\mathrm{c} = 4$\,mm) is used to improve field uniformity. FEA results (dashed and dotted curves in Fig. \ref{fig01}(c)) show that the compensation expands the uniform field range by about 34\%, from 23\,mm to 30.8\,mm, maintaining a relative variation $\Delta B_\mathrm{r}/ B_\mathrm{r} < 2 \times 10^{-3}$, where $B_\mathrm{r}$ is the radial field and $\Delta B_\mathrm{r}$ its variation. The splitting surface at $z = 11$\,mm allows a levitation distance of over 40\,mm when open and generates approximately 92\,N attractive force when closed.

As experimentally observed in Fig.~\ref{fig01}(c), the $B_\mathrm{r}(z)$ profile exhibits an asymmetry, quantified by the difference between the upper and lower peak values of the magnetic flux density, 
\begin{equation}
    \Delta B_\mathrm{r} = B_\mathrm{up} - B_\mathrm{dn},
\end{equation}
where \( B_\mathrm{up} \) and \( B_\mathrm{dn} \) denote {peak values with the maximum and minimum $z$ coordination}, respectively. Note that here \( B_\mathrm{dn} \) is chosen as the reference, as the profile shape around this point $z=z_\mathrm{dn}$ is not considerably affected by the splitting plane. 
For the measured profile, this asymmetry amounts to approximately $\Delta B_\mathrm{r}\approx$5.5\,mT.  

\begin{figure}[tp!]
\centering
\includegraphics[width=0.32\textwidth]{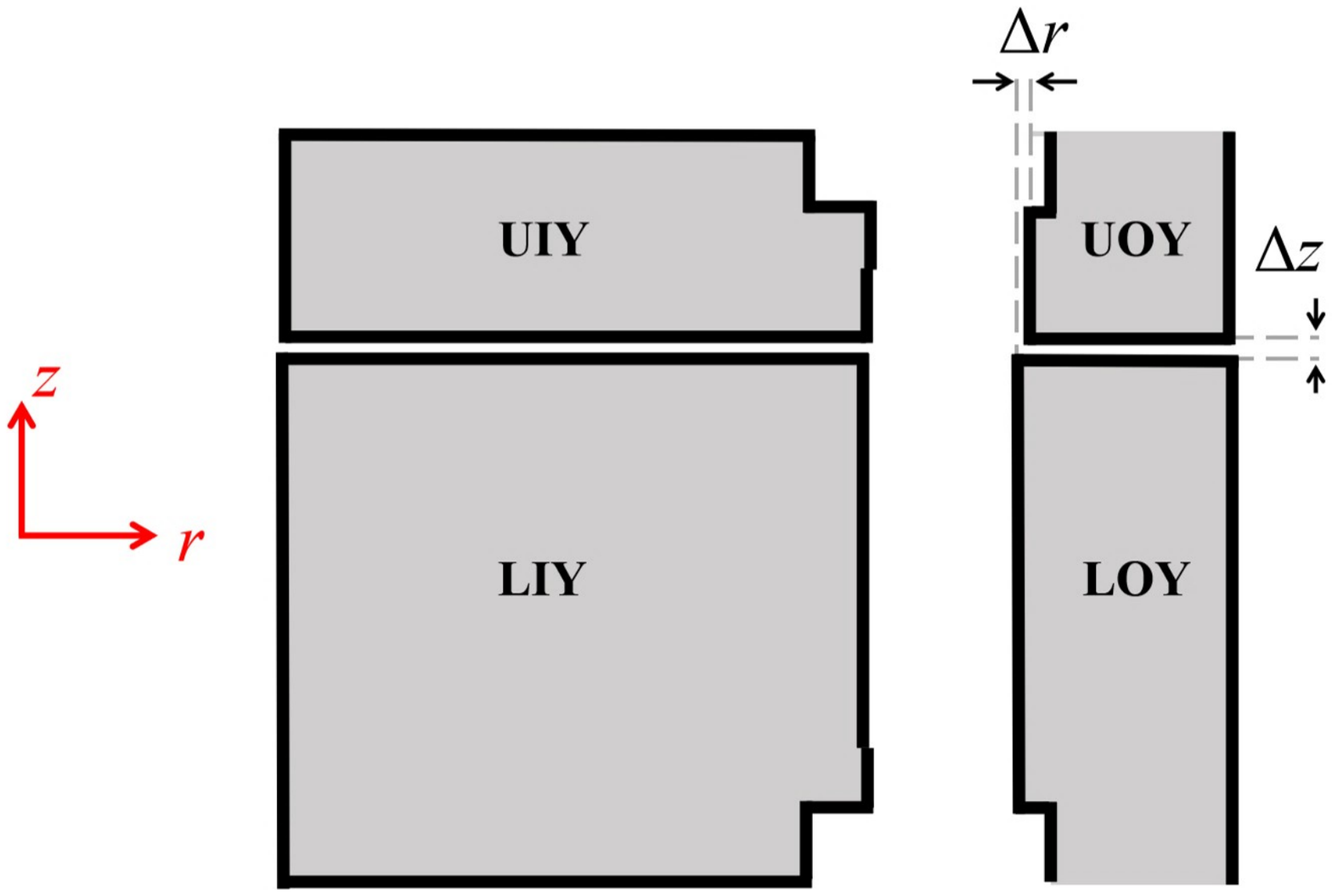}
 \caption{Partial cross-sectional view of the magnet assembly showing geometric parameters $\Delta z$ and $\Delta r$. 
The splitting air gap $\Delta z$ and inner radius adjustment $\Delta r$ are indicated by dashed lines. 
Component labels identify the upper inner yoke (UIY), upper outer yoke (UOY), lower inner yoke (LIY), and lower outer yoke (LOY).}
\label{fig03}
\end{figure}

\begin{figure*}[tp!]
\centering
\includegraphics[width=0.8\textwidth]{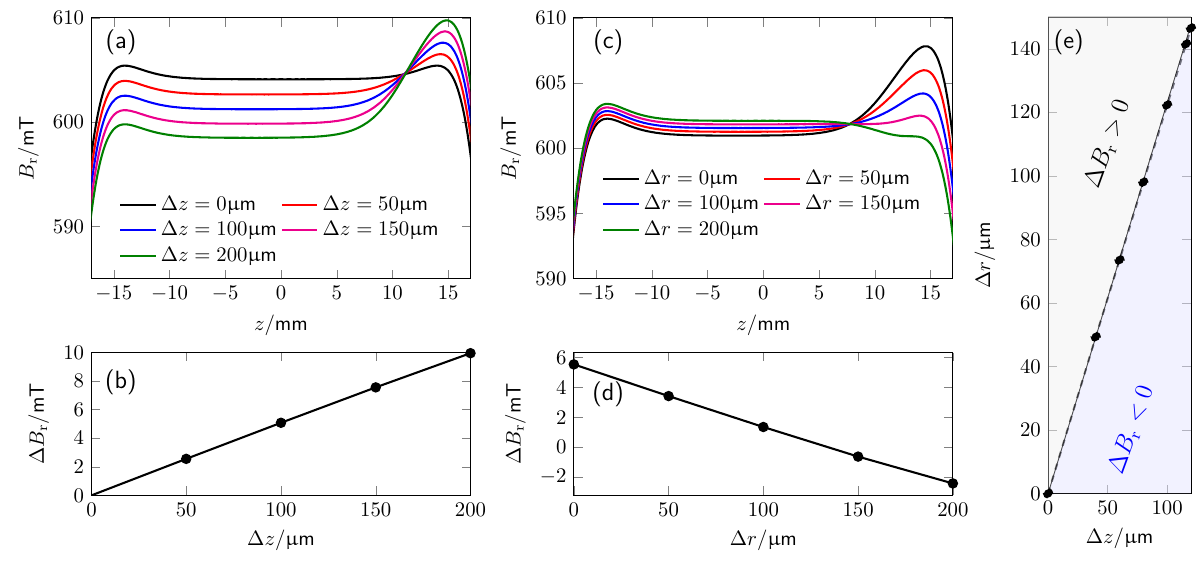}
 \caption{Influence of key geometric parameters on magnetic field asymmetry. 
(a) Asymmetry $\Delta B_{\mathrm{r}}$ increases with splitting air gap height $\Delta z$. 
(b) Linear dependence $\Delta B_{\mathrm{r}}(\Delta z) = \alpha \Delta z$ is quantified, with $\alpha = 50.13$\,mT/mm. 
(c) Increasing upper yoke radius adjustment $\Delta r$ reduces the peak field $B_\mathrm{up}$.
(d) Corresponding linear relation $\Delta B_{\mathrm{r}}(\Delta r) = \beta \Delta r+\gamma$ is shown, where $\beta = -40.02$\,mT/mm and $\gamma=5.45$\,mT. 
(e) Combined influence of $\Delta z$ and $\Delta r$ on $\Delta B_\mathrm{r}$ is presented. Magnetic symmetry ($\Delta B_\mathrm{r}= 0$) defines a linear compensation line between $\Delta z$ and $\Delta r$.
}
\label{fig02}
\end{figure*}

To investigate the influence of the splitting air gap {height, \(\Delta z\) as indicated in Fig. \ref{fig03},} on \(\Delta B_\mathrm{r}\), FEA simulations were performed for varying   
\(\Delta z\). Only permanent magnets and ferromagnetic yoke components are modeled in the simulation, excluding auxiliary elements. The results, presented in Fig.~\ref{fig02}(a), reveal that the asymmetry becomes more pronounced with increasing \(\Delta z\). Furthermore, Fig.~\ref{fig02}(b) demonstrates a linear relationship between \(\Delta B_\mathrm{r}\) and \(\Delta z\), which can be expressed as  
\begin{equation}
    \Delta B_\mathrm{r}(\Delta z) = \alpha \Delta z,
    \label{eqBz}
\end{equation}
where \(\alpha\) is the linear coefficient. In this case, \(\alpha\) was determined to be 50.13\,mT/mm. Based on this relationship, the measured asymmetry of \(\Delta B_\mathrm{r} = 5.5\)\,mT corresponds to an equivalent splitting air gap of \(\Delta z = 109.58\,\upmu\)m.

This asymmetry originates from an unavoidable assembly-induced splitting air gap, which increases the magnetic field in the upper region, especially above the splitting plane at \( z = 11 \, \text{mm} \). As established~\cite{Li_2022_irony}, the radial magnetic field in the main air gap approximately follows a \( 1/r \) distribution, so that enlarging the air gap reduces the flux density. For example, increasing the inner radius of the upper outer yoke by \( \Delta r \), while keeping the inner and lower outer yoke dimensions fixed as in Fig.~\ref{fig03}, can weaken the field above the splitting plane. This modification provides a potential method to reduce the profile asymmetry.
To investigate the effect of adjusting the upper outer yoke radius ($\Delta r$) on the magnetic profile, we also performed FEA simulations. These simulations maintained a constant splitting air gap height of $\Delta z = 109.58\,\upmu\text{m}$ (determined from experimental measurements) while varying $\Delta r$ from $0\,\upmu\text{m}$ to $200\,\upmu\text{m}$. 
Fig.~\ref{fig02}(c) presents the resulting magnetic field distribution, showing a progressive decrease in the upper peak field ($B_\mathrm{up}$) with increasing $\Delta r$. As demonstrated in Fig.~\ref{fig02}(d), the asymmetry $\Delta B$ follows a linear relationship with $\Delta r$, written as
\begin{equation}
\Delta B_\mathrm{r}(\Delta r) = \beta \Delta r + \gamma,
\label{eqBr}
\end{equation}
where $\beta$ is the linear coefficient, in this case found to be $-40.02\,\text{mT/mm}$, and $\gamma$ denotes a constant offset. The negative value of $\beta$ confirms that increasing the yoke radius reduces the upper field peak value $B_\mathrm{up}$.

Combining (\ref{eqBz}) and (\ref{eqBr}) yields the composite relationship,
\begin{equation}
\Delta B_\mathrm{r}(\Delta z, \Delta r) = \alpha \Delta z + \beta \Delta r.
\label{eqBzr}
\end{equation}
Note that when $\Delta z = \Delta r = 0$, the magnetic profile becomes perfectly symmetrical ($\Delta B_\mathrm{r} = 0$), which implies $\gamma = 0$ in (\ref{eqBr}). Setting $\Delta B_\mathrm{r} = 0$ gives the required radius adjustment for the upper yoke,
\begin{equation}
\Delta r = -\frac{\alpha}{\beta} \Delta z.
\end{equation}
For the experimental case shown in Fig.~\ref{fig01}(c), a radius adjustment of $\Delta r = 136.27\,\upmu\text{m}$ can well eliminate the profile asymmetry. More generally, for any given $\Delta z$, there exists an optimal $\Delta r$ that satisfies $\Delta B_\mathrm{r} = 0$. Fig.~\ref{fig02}(e) illustrates a two-dimensional relationship between $\Delta z$ and $\Delta r$ with different $\Delta B_\mathrm{r}$ values. The linear curve represents the condition $\Delta B_\mathrm{r} = 0$, with regions above and below the curve corresponding to positive and negative $\Delta B_\mathrm{r}$ values, respectively.

In practice, precisely adjusting $\Delta r$ to compensate for the profile asymmetry based on experimental determination of $\Delta z$ presents significant challenges. The required $\Delta r$ values typically range from tens to hundreds of micrometers, making both implementation and high-resolution measurement difficult.
A more robust operational approach, which involves two key steps, is proposed:
First, intentionally set $\Delta r$ such that $|\beta \Delta r| > |\alpha \Delta z|$, creating a controlled asymmetry where the upper field region becomes weaker than the lower region ($\Delta B_\mathrm{r} < 0$).
Then, manually increase $\Delta z$ through fine splitting air gap adjustments (e.g., via shimming\cite{NISTmag}) until satisfying the compensation condition
$\alpha \Delta z + \beta \Delta r = 0$.
This two-step yoke compensation method offers distinct advantages: $\Delta z$ can be continuously adjusted with greater ease compared to $\Delta r$, and an optimal compensation condition can always be achieved through this iterative process.


\section{Experimental results}
\label{sec03}

\begin{figure}[tp!]
\centering
\includegraphics[width=0.5\textwidth]{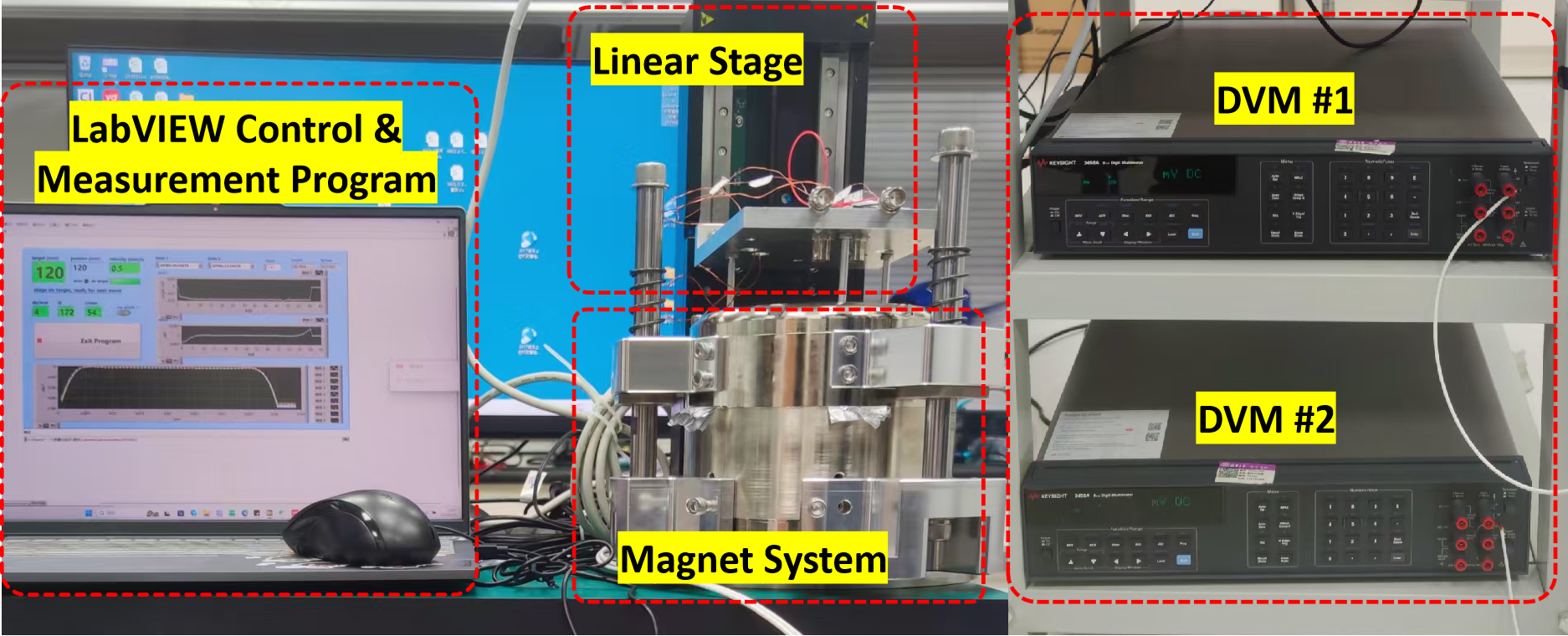}
 \caption{A photo of the experimental setup. A linear stage, PI M-413.2DG, is used to move the gradient coil up and down. Two DVMs, \#1 and \#2, are used to measure {the induced voltage of one of the two coils and the differential voltage of two coils connected in opposite directions, respectively.} }
\label{Laboratory Bench}
\end{figure}

To validate the proposed two-step yoke compensation method, experiments were performed on the magnet system in Fig.~\ref{fig01}(b). The air-gap magnetic field was measured using a gradient coil~\cite{NISTmag}, with the setup shown in Fig.~\ref{Laboratory Bench}. Magnet parameters matched those used in simulations (Section~\ref{sec02}).
The gradient coil comprises two identical coils vertically separated by $d=4$\,mm, each with 172 turns and a main radius of 54.5\,mm. The magnetic field distribution is measured via the gradient coil configuration. During vertical motion at velocity $v$, the following relationship is established \cite{NISTmag}:
\begin{equation}
\frac{V_1(z) - V_2(z)}{V_1(z)} = \frac{B_\mathrm{r}(z) - B_\mathrm{r}(z - d)}{B_\mathrm{r}(z)} \approx \frac{d}{B_\mathrm{r}(z)} \frac{\mathrm{d}B_\mathrm{r}(z)}{\mathrm{d}z},
\end{equation}
where $V_1$ and $V_2$ denote induced voltages in the respective coils. The magnetic field distribution $B_\mathrm{r}(z)$ is derived as
\begin{equation}
B_\mathrm{r}(z) = \frac{\bar{B_\mathrm{r}}}{\bar{V}_1 d} \int ( V_1(z) - V_2(z) ) \mathrm{d}z + B_0,
\end{equation}
where $B_0$ is an integration constant satisfying $B_\mathrm{r}(0) = \bar{B_\mathrm{r}}$\cite{NISTmag}. Mounted on a linear stage (PI M-413.2DG), the gradient coil was connected to two digital voltmeters (Keysight 3458A) to measure both the induced voltage $\bar{V}_1$ and the differential voltage between the coils $V_1- V_2$. 

\begin{figure}[tp!]
\centering
\includegraphics[width=0.4\textwidth]{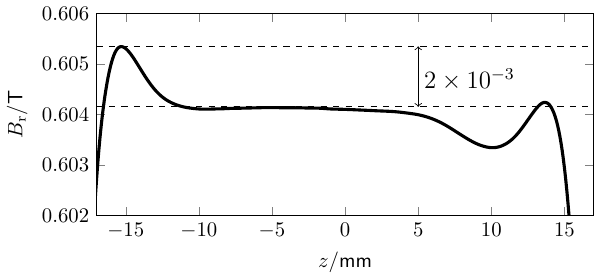}
 \caption{The measured air-gap flux density following the adjustment of the upper outer yoke's inner diameter increment.
}
\label{newB0}
\end{figure}

After a slight machining enlargement of the inner radius of the upper outer yoke, the resulting magnetic field distribution is shown in Fig.~\ref{newB0}. In this case, it is observed that $|\beta \Delta r| > |\alpha \Delta z|$ and $B_{\mathrm{up}} < B_{\mathrm{dn}}$. The introduction of $\Delta r$, as predicted by the FEA simulation, effectively reduces the magnetic flux density in the upper region of the profile.

\begin{figure}[tp!]
\centering
\includegraphics[width=0.4\textwidth]{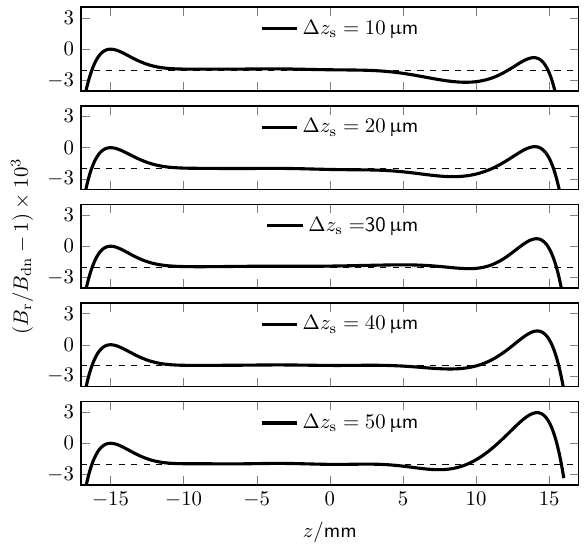}
 \caption{Measured magnetic field distributions at varying shim thicknesses ($\Delta z_\mathrm{s} = 10\,\upmu$m to $50\,\upmu$m) using shims inserted at the splitting plane, where $\Delta z_\mathrm{s}$ means thickness of the shim.}
\label{Experimental Result}
\end{figure}

Next, thin aluminum shims were inserted at the splitting plane, and the magnetic field profile was measured using the setup described above. Fig.~\ref{Experimental Result} shows the magnetic field distributions for shim thicknesses $\Delta z_\mathrm{s}$ from 10 to 50\,$\upmu$m in 10\,$\upmu$m steps. As $\Delta z_\mathrm{s}$ increases, $B_\mathrm{up}$ rises consistently with FEA predictions. At $\Delta z_\mathrm{s} = 30$\,$\upmu$m, the profile becomes effectively flat, satisfying $|\beta \Delta r| \approx |\alpha \Delta z|$. Compared to the initial uncompensated case (Fig.~\ref{fig01}(c)), the yoke modification restores field flatness with good symmetry. The uniform field range reaches about 30\,mm, matching the ideal design. These results confirm that sequential adjustment of $\Delta r$ and $\Delta z$ effectively eliminates profile asymmetry induced by the splitting air gap.

\section{Conclusion}
\label{sec04}

The magnetic profile asymmetry in openable BIPM-type Kibble balance magnets, caused by structural splitting air gaps, leads to a localized field increase near the splitting plane. To address this, a two-step yoke compensation method is proposed, combining radial adjustment of the upper yoke ($\Delta r$) with fine-tuning of the splitting gap height ($\Delta z$). Theoretical analysis shows that the asymmetry $\Delta B_\mathrm{r}$ depends linearly on both parameters:$\Delta B_\mathrm{r}(\Delta z, \Delta r) = \alpha \Delta z + \beta \Delta r,$where $\alpha > 0$ and $\beta < 0$. For a given $\Delta B_\mathrm{r}$ induced by $\Delta z$, an optimal $\Delta r$ can restore symmetry. As $\Delta r$ adjustment is inflexible, a practical compensation procedure is adopted: first, $\Delta r$ is slightly enlarged by machining to create $\Delta B_\mathrm{r} < 0$; then, $\Delta z$ is increased to raise $\Delta B_\mathrm{r}$ until symmetry is recovered. This approach allows flexible fine-tuning via $\Delta z$. Experimental validation demonstrates successful elimination of a $\Delta B_\mathrm{r} = 5.5$\,mT asymmetry, improving $Bl$ measurement accuracy. The method offers a reproducible strategy for enhancing magnet performance in systems requiring open/close operation.


\begin{thebibliography}{10}


\bibitem{kibble1976measurement}
B.~P.~Kibble, ``A measurement of the gyromagnetic ratio of the proton by the strong field method,'' \emph{Atomic masses and fundamental constants 5}, pp. 545--551, 1976.

\bibitem{haddad2016bridging}
D.~Haddad, F.~Seifert, L.~S. Chao \textit{et al.}, ``Bridging classical and quantum mechanics,'' \emph{Metrologia}, vol.~53, no.~5, p. A83, 2016.

\bibitem{Robinson_2016}
I.~A. Robinson and S.~Schlamminger, ``The watt or {K}ibble balance: a technique for implementing the new {SI} definition of the unit of mass,'' \emph{Metrologia}, vol.~53, no.~5, p. A46, Sep 2016.

\bibitem{Li_2022}
S.~Li and S.~Schlamminger, ``The irony of the magnet system for Kibble balances—a review,'' \emph{Metrologia}, vol.~59, no.~2, p. 022001, Mar 2022.



\bibitem{Wood_2017}
B.~M. Wood, C.~A. Sanchez, R.~G. Green \textit{et al.}, ``A summary of the {P}lanck constant determinations using the {NRC} {K}ibble balance,'' \emph{Metrologia}, vol.~54, no.~3, p. 399, May 2017.

\bibitem{NIST2017}
D.~Haddad, F.~Seifert, L.~S. Chao \textit{et al.}, ``{M}easurement of the {P}lanck constant at the {N}ational {I}nstitute of {S}tandards and {T}echnology from 2015 to 2017,'' \emph{Metrologia}, vol.~54, no.~5, p. 633, Jul 2017.


\bibitem{METAS2022}
A.~Eichenberger, H.~Baumann, A.~Mortara \textit{et al.},``{F}irst realisation of the kilogram with the {METAS} {K}ibble balance,'' \emph{Metrologia}, vol.~59, no.~2, p. 025008, Mar 2022.

\bibitem{Fang_2020BIPM}
H.~Fang, F.~Bielsa, S.~Li \textit{et al.}, ``{T}he {BIPM} {K}ibble balance for realizing the kilogram definition,'' \emph{Metrologia}, vol.~57, no.~4, p. 045009, Jul 2020. 

\bibitem{NIM2023}
Z.~Li, Y.~Bai, Y.~Wang \textit{et al.}, ``{I}mprovements of the {NIM}-2 {J}oule balance since 2020,'' \emph{IEEE Transactions on Instrumentation and Measurement}, vol.~72, pp. 1--7, 2023. 

\bibitem{KRISS2020}
D.~Kim, M.~Kim, M.~Seo \textit{et al.}, ``{R}ealization of the kilogram using the {KRISS} {K}ibble balance,'' \emph{Metrologia}, vol.~57, no.~5, p. 055006, Sep 2020.


\bibitem{UME2023}
H.~Ahmedov, R.~Orhan, and B.~Korutlu, ``{UME} {K}ibble balance operating in air,'' \emph{Metrologia}, vol.~60, no.~1, p. 015003, Dec 2022. 

\bibitem{MSL2020}
Y.~Fung, M.~Clarkson, and F.~Messerli, ``{A}lignment in the {MSL} {K}ibble balance,'' in \emph{2020 Conference on Precision Electromagnetic Measurements (CPEM)}, 2020, pp. 1--2.

\bibitem{NPLtabletop2024}

I.~A. Robinson, A.~E. Karsten, T.~Mametja \textit{et al.},``Progress on the {NPL}, {NMISA}, {RISE} {K}ibble {B}alance {C}ollaboration,'' in \emph{2024 Conference on Precision Electromagnetic Measurements (CPEM)}, 2024, pp. 1--2. 

\bibitem{LNE2025}
M.~Thomas, P.~Espel, K.~Dougdag \textit{et al.},``Mass measurement noise improvement of the LNE Kibble balance, from 2017 to 2024,'' \emph{IEEE Transactions on Instrumentation and Measurement}, vol.~74, pp. 1--7, 2025.

\bibitem{Rothleitner_2018}

C.~Rothleitner, J.~Schleichert, N.~Rogge \textit{et al.}, ``The Planck-balance—using a fixed value of the Planck constant to calibrate E1/E2-weights,'' \emph{Measurement Science and Technology}, vol.~29, no.~7, p. 074003, May 2018. 

\bibitem{THU2025}
S. Li, Y. Ma, K. Ma \textit{et al.}, ``Updates on the Tsinghua Tabletop Kibble Balance,'' \emph{IEEE Transactions on Instrumentation and Measurement}, vol.~74, pp. 1--9, 2025.




\bibitem{BIPMmag}
A.~Picard, M.~Stock, H.~Fang \textit{et al.}, ``The BIPM watt balance,'' \emph{IEEE Transactions on Instrumentation and Measurement}, vol.~56, no.~2, pp. 538--542, 2007.

\bibitem{NISTmag}
F.~Seifert, A.~Panna, S.~Li \textit{et al.}, ``Construction, measurement, shimming, and performance of the NIST-4 magnet system,'' \emph{IEEE Transactions on Instrumentation and Measurement}, vol.~63, no.~12, pp. 3027--3038, 2014.

\bibitem{NIMmag}

Z.~Li, Z.~Zhang, Y.~Lu \textit{et al.}, ``The first determination of the {P}lanck constant with the joule balance {NIM-2},'' \emph{Metrologia}, vol.~54, no.~5, p. 763, Sep 2017. 


\bibitem{METAS_Baumann_2013}
H.~Baumann, A.~Eichenberger, F.~Cosandier \textit{et al.}, ``{D}esign of the new {METAS} watt balance experiment {M}ark {II},'' \emph{Metrologia}, vol.~50, no.~3, p. 235, May 2013. 


\bibitem{Kim_KRISS_2014}
D.~Kim, B.-C. Woo, K.-C. Lee \textit{et al.}, ``Design of the KRISS watt balance,'' \emph{Metrologia}, vol.~51, no.~2, p. S96, Mar 2014. 


\bibitem{THUmag}
Y.~Ma, N.~Li, W.~Liu \textit{et al.}, ``A compact magnet system for the Tsinghua tabletop Kibble balance,'' \emph{IEEE Transactions on Instrumentation and Measurement}, vol.~74, pp. 1--13, 2025.

\bibitem{UME_Ahmedov_2018}

H.~Ahmedov, N.~B. Aşkın, B.~Korutlu \textit{et al.}, ``Preliminary Planck constant measurements via UME oscillating magnet Kibble balance,'' \emph{Metrologia}, vol.~55, no.~3, p. 326, Apr 2018. 








\bibitem{9042232}
S.~Li, S.~Schlamminger, and Q.~Wang, ``A simple improvement for permanent magnet systems for Kibble balances: More flat field at almost no cost,'' \emph{IEEE Transactions on Instrumentation and Measurement}, vol.~69, no.~10, pp. 7752--7760, 2020.

\bibitem{NIST4magdesign}
S. Schlamminger, ``Design of the permanent-magnet system for NIST-4,'' \emph{IEEE Transactions on Instrumentation and Measurement}, vol.~62, no.~6, pp. 1524-1530, 2012.

\bibitem{Magneticerror}
S. Li, S. Schlamminger, ``Magnetic uncertainties for compact Kibble balances: An investigation,'' \emph{IEEE Transactions on Instrumentation and Measurement}, vo.~71, 1502409, July 2022.

\bibitem{thermalerror}
W. Liu, S. Schlamminger, S. Li, ``Precision control of resistive power in Kibble balance coils: An advanced method for minimizing temperature-related magnetic errors,'' \emph{Metrologia}, vol.~62, no.~3,  035009, 2025.

\bibitem{flatprofile}
S. Li, S. Schlamminger, D. Haddad, et al, ``Coil motion effects in watt balances: a theoretical check,'' \emph{Metrologia}, vol. 53, no. 2, pp. 817, 2016.

\bibitem{BNMmag}
P. Gournay G. Genevès, F. Alves, \textit{et al.}, ``Magnetic circuit design for the BNM watt balance experiment,'' \emph{IEEE Transactions on instrumentation and Measurement}, vol. 54, no. 2, pp.742-745, 2005.

\bibitem{METASmag}
A. L. Eichenberger, J. Butty, B. Jeanneret \textit{et al.}, ``A New Magnet Design for the METAS Watt Balance,'' \emph{2004 Conference on Precision Electromagnetic Measurements}, pp. 56--57, 2004. 

\end{thebibliography}

\end{document}